\documentclass[aps,prl,preprint,superscriptaddress]{revtex4}

 \usepackage{amsmath,bm}
 \usepackage{mathrsfs}
 \usepackage{amsfonts}
 \usepackage{graphicx}
 \usepackage{ulem}
 \usepackage{setspace} 
 \usepackage{graphicx}
 \usepackage{epstopdf}
 \usepackage{dcolumn}
 \usepackage{amsmath}
 \usepackage{epsfig}
 \usepackage{indentfirst}
 \usepackage{psfrag}
 \usepackage{subfigure}
 \usepackage{amssymb}
 \usepackage{color}

 \usepackage{graphicx}
 \usepackage{dcolumn}
 \usepackage{bm}
 \usepackage{natbib}

\usepackage{physics}
\usepackage{dcolumn}
\usepackage{bm}
\usepackage{hyperref}
\usepackage{natbib}
\makeatletter
\newcommand\colorsout[1]{\bgroup \markoverwith{\textcolor{#1}{\rule[0.5ex]{2pt}{0.4pt}}}\ULon}

\makeatother

\begin{document}


\title{Orbital-selective spin excitation of a magnetic porphyrin}

\author{C. Rubio-Verd\'u }
  \affiliation{CIC nanoGUNE, 20018 Donostia-San Sebasti\'an, Spain}
 
\author{A. Sarasola }
\affiliation{Donostia International Physics Center (DIPC), 20018 Donostia-San Sebasti\'an, Spain}
 \affiliation{Faculty of Engineering, Gipuzkoa, 20018 Donostia-San Sebasti\'an, Spain}

 \author{D. -J. Choi}
 \affiliation{CIC nanoGUNE, 20018 Donostia-San Sebasti\'an, Spain}
 \affiliation{Donostia International Physics Center (DIPC), 20018 Donostia-San Sebasti\'an, Spain}
\affiliation{Centro de F\'isica de Materiales (CFM), 20018 Donostia-San Sebasti\'an,  Spain}

\author{Z. Majzik}
 \affiliation{CIC nanoGUNE, 20018 Donostia-San Sebasti\'an, Spain}

\author{R. Ebeling}
 \affiliation{CIC nanoGUNE, 20018 Donostia-San Sebasti\'an, Spain}

\author{M. R. Calvo}
 \affiliation{CIC nanoGUNE, 20018 Donostia-San Sebasti\'an, Spain}
\affiliation{Ikerbasque, Basque Foundation for Science, Bilbao, Spain}

\author{M. M. Ugeda}
 \affiliation{CIC nanoGUNE, 20018 Donostia-San Sebasti\'an, Spain}
 \affiliation{Donostia International Physics Center (DIPC), 20018 Donostia-San Sebasti\'an, Spain}
\affiliation{Ikerbasque, Basque Foundation for Science, Bilbao, Spain}
\affiliation{Centro de F\'isica de Materiales (CFM), 20018 Donostia-San Sebasti\'an,  Spain}

\author{A. Garcia-Lekue}
\affiliation{Donostia International Physics Center (DIPC), 20018 Donostia-San Sebasti\'an, Spain}
\affiliation{Ikerbasque, Basque Foundation for Science, Bilbao, Spain}
 
\author{D. S\'anchez-Portal}
\affiliation{Donostia International Physics Center (DIPC), 20018 Donostia-San Sebasti\'an, Spain}
\affiliation{Centro de F\'isica de Materiales (CFM), 20018 Donostia-San Sebasti\'an,  Spain}

\author{J. I. Pascual}
 \affiliation{CIC nanoGUNE, 20018 Donostia-San Sebasti\'an, Spain}
\affiliation{Ikerbasque, Basque Foundation for Science, Bilbao, Spain}

\begin{abstract}

Scattering of electrons by localized spins is the ultimate process enabling electrical detection and control of the magnetic state of a spin-doped material. At the molecular scale, this scattering is mediated by the electronic orbitals hosting the spin. Here we report the selective excitation of a molecular spin by electrons tunneling through different molecular orbitals. Spatially-resolved tunneling spectra on iron porphyrins on Au(111) reveal that the inelastic spin excitation extends beyond the iron site. The inelastic features also change shape and symmetry along the molecule. Combining DFT simulations with a phenomenological scattering model, we show that the extension and lineshape variations of the inelastic signal are due to excitation pathways assisted by different frontier orbitals, each of them with a different degree of hybridization with the surface. By selecting the intramolecular site for electron injection, the relative weight of iron and pyrrole orbitals in the tunneling process is modified. In this way, the spin excitation mechanism, reflected by its spectral lineshape, changes depending on the degree of localization and energy alignment of the chosen molecular orbital. 
\end{abstract}


\maketitle

Organic spintronics has accomplished such a level of development in the recent years that has been labeled as an emerging technology \cite{joachim_electronics_2000,wolf_spintronics:_2001,bogani_molecular_2008, schmaus_giant_2011,dediu_spin_2009,Sun2017}. Its realization depends on the efficient interaction of electron currents with spins of a magnetic organic material.  The spin density of an organometallic  molecule is generally localized at the spin-doping metal ion, but it can be accessed through its organic ligand environment.  For instance, the molecular spin can be manipulated by  distortions of the molecular structure  \cite{Iancu2006,parks_mechanical_2010,Komeda2011,heinrich_change_2013,heinrich_tuning_2015,auwarter_porphyrins_2015,Kuang2017}, or by tuning the molecular interaction with the local environment \cite{zhao_controlling_2005,Iancu2006b,zhao_kondo_2008,Franke2011b,Strozecka2012,Oberg2013}.

In most organometallic species, the spin of the metallic ion is distributed among several spin-polarized atomic orbitals. The organic ligands around cast an anisotropic field that breaks the orbital degeneracy and induces a preferential spin orientation. The split spin-hosting orbitals extend and mix differently with molecular states and with the surroundings, behaving as different electron channels for scattering with the spin.   At the single molecule scale, the existence of several electron-spin scattering channels competing in a molecule has been usually identified indirectly by phenomena such as the partial  Kondo screening of the spin (underscreened Kondo \cite{parks_mechanical_2010,Strozecka2012}), the spatial extension of Kondo spectral features \cite{perera_spatially_2010}, or by the effect of the molecular spin on superconducting quasiparticles \cite{Heinrich2013}.

Here, we show that a molecular spin can be selectively excited by injecting electrons into two different orbital channels. We investigate the spin-excitation of Fe-tetraphenyl porphyrin (FeTPP) molecules adsorbed on a Au(111) surface by Scanning Tunneling Spectroscopy (STS). The spin $S =1$ of FeTPP can be excited either by electron tunneling over the Fe ion or over two of the four molecular pyrrole groups. Interestingly, the inelastic spin excitation features show asymmetric components with opposite sign on each site. Asymmetry in the $dI/dV$ spectra of magnetic systems has been tipically attributed to Fano interference in a Kondo-screened spin \cite{Iancu2006,wang_intramolecularly_2015}. However, here Kondo screening is absent and we show that the spectral asymmetry is due to the spin-polarized orbital lying out of the particle-hole symmetry point. Then, each spectral shape reveals the alignment and hybridization of the molecular states hosting the molecular spin. With the support of  Density Functional Theory (DFT) simulations, we provide an atomistic picture of each tunneling  channel based on  distinct spin-polarized molecular states with strong Fe character and different distribution within the molecule.

\section*{Results}

\begin{figure}[!ht]
	\includegraphics[width=0.4\textwidth]{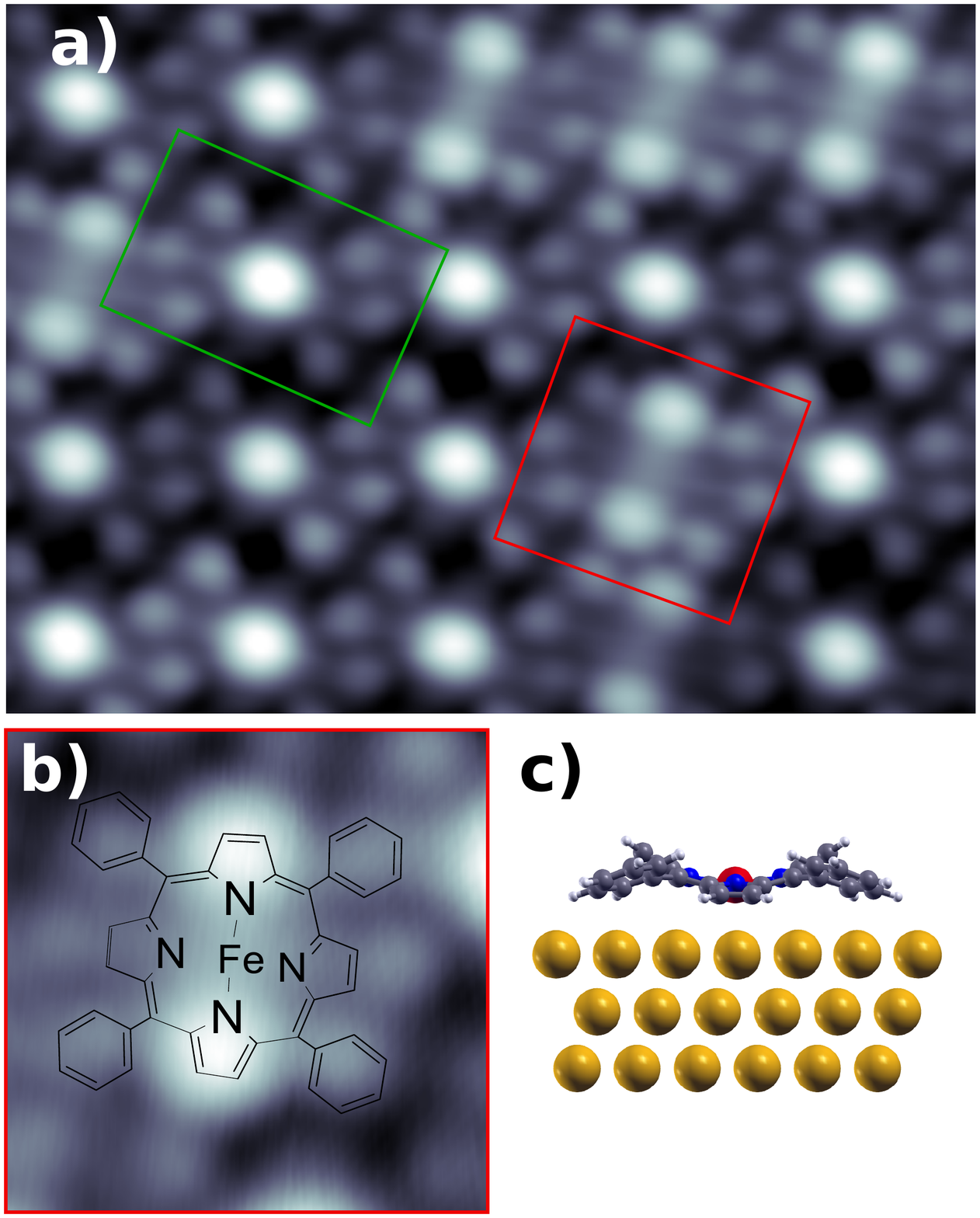}
	\caption{\label{Fig1} (a) Topographic STM image (7 $\times$ 5 nm$^2$) of a close-packed island of FeTPP (red box) and FeTPP-Cl (green box) molecules on Au(111). Imaging conditions: $ V=0.25$ V, $I=10$ pA. (b) Zoom of a single FeTPP molecule and molecular structure superimposed. (c) Cross section of the DFT relaxed structure of the FeTPP on the Au(111) surface (3 gold layers) showing the saddle conformation acquired upon adsorption. }
\end{figure}

STM images show that the porphyrin molecules arrange in close-packed islands on the Au(111) surface (Fig. \ref{Fig1}(a)) and appear with a protrusion at the center (green box in Fig. \ref{Fig1}(a)) that is attributed to the Cl ligand of the intact FeTPP-Cl molecule. A fraction of the molecules exhibits instead two lobes. They correspond to the dechlorinated specie FeTPP (Fig. \ref{Fig1}(b)). FeTPP molecules can be controllably obtained by removing the Cl ligand from the FeTPP-Cl molecules using tunneling electrons (see Supplemental  Material (SM) \cite{SOM}) or by annealing the substrate \cite{heinrich_change_2013}. The dechlorination process changes the Fe oxidation state from Fe$^{+3}$ to Fe$^{+2}$ and decreases the total spin from $S=3/2$ to $S=1$ \cite{gopakumar_transfer_2012,heinrich_change_2013,heinrich_tuning_2015}.

\begin{figure*}[t] 
	\includegraphics[width=1\textwidth]{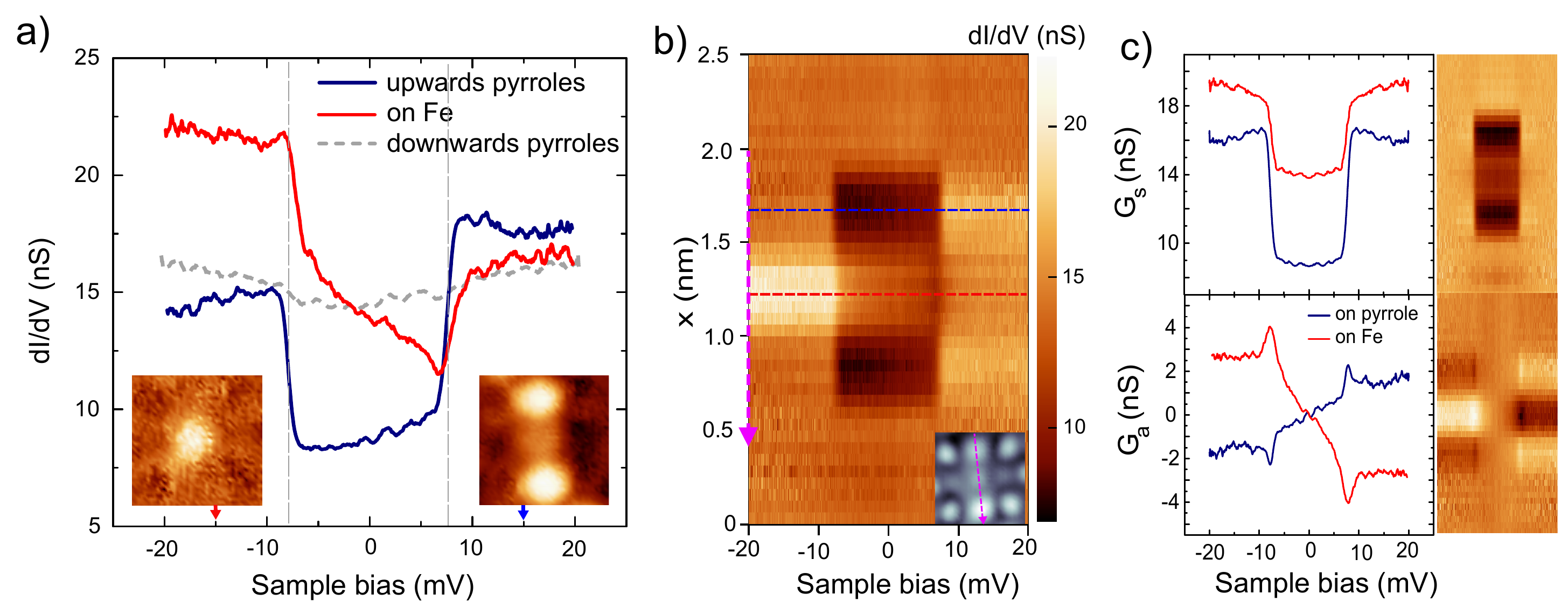}
	\caption{\label{Fig2}  (a) Characteristic $dI/dV$ spectra of FeTPP measured over one of the two 
upward  pyrrole groups (blue) and over the central Fe ion (red). As a 
reference, we add a spectrum acquired over downward pyrroles with the 
same tip (dashed grey), showing the complete absence of features in this 
range. (Setpoint: $V=20$ mV, $I=300$ pA. Lock-in frequency 938 Hz, modulation \mbox{50 $\mu$V {\it{rms}}}). The insets  are 1.7$\times$1.7 nm$^2$ constant current $dI/dV$ maps at V=-15 mV and V=+15 mV of the topography image shown as inset in (b).  (b) Stacking plot of point $dI/dV$ spectra (40 curves, 2.5 nm) along the axis of a FeTPP molecule (as shown in the inset). (c) Symmetric ($G_{s} = \frac{1}{2} ( G_{V>0} + G_{V<0} )$) and antisymmetric ($G_{a} = \frac{1}{2} ( G_{V>0} - G_{V<0})$) components of the $dI/dV$ spectra of (a) and their flatten spatial distribution along the axis of the FeTPP molecule, as in (b).  }
\end{figure*}

STS measurements over the molecules reveal steps in conductance at symmetric bias values (Fig. \ref{Fig2}(a)), associated to the onset of  inelastic tunneling \cite{stipe_single-molecule_1998,a._j._heinrich_single-atom_2004}.  While the steps on the FeTPP-Cl molecules appear at $V_{S}=\pm1.7$ mV (shown in SM \cite{SOM}), the removal of the Cl ligand rises the excitation energy to $V_{S}=\pm7.4$ mV. 
In both cases the inelastic spectra shows a dependence with the applied magnetic field (see SM \cite{SOM}), which agrees with changes of the spin-multiplet of Fe-porphyrin moieties \cite{heinrich_change_2013-1,heinrich_protection_2013,heinrich_tuning_2015}. We, thus, associated these steps to spin excitations induced by tunneling electrons, and discard any possible vibrational origin \cite{vib_chen}.

We find a striking spatial distribution of the spectroscopic features of the FeTPP species (see Fig. \ref{Fig2}(a)). While the energy position of the inelastic conductance steps remains the same ($V_{S}=\pm7.4 \pm 0.5$ mV) all over the molecule, their symmetry varies as we move off-center towards the brighter pyrroles. A stacking plot of point $dI/dV$ spectra across the FeTPP  molecule (Fig. \ref{Fig2}(b)) shows inelastic steps with a rather symmetric lineshape over the pyrroles and with strong antisymmetric character over the Fe site. It is remarkable that the energy positions of the conductance steps remain nevertheless constant, indicating that the same excitation is the origin for the inelastic features across the molecule. The spin excitation is only observed along the axis formed by the brighter pyrroles (marked in the inset in Fig. \ref{Fig2}(b)), and is absent over the two other pyrroles (grey dashed line in Fig. \ref{Fig2}(a)).

In a first approximation, the spin excitation energy is determined by the magnetic anisotropy of the molecule on the surface, which can be modeled with a phenomenological spin Hamiltonian  \cite{gatteschi_molecular_2006}
$ \hat{H_s} =  D \hat{S_z^2} + E(\hat{S_x^2} - \hat{S_y^2} ) $, where $\hat{S_x}, \hat{S_y}, \hat{S_z}$ 
are the three components of the spin operator, and $D$ and $E$ the axial and transverse anisotropy parameters, respectively.
We find that the inelastic step   splits by applying a magnetic field perpendicular to the surface, in accordance with Zeeman shifts of a S=1 multiplet (see Fig. S3 in SM \cite{SOM}).  
Hence, the excitation at V$_{s}=\pm 7.4$ mV corresponds to a transition from the $\ket{m_S= 0}$ ground state to the $\ket{m_S=\pm 1}$ multiplet, and the excitation energy is the axial anisotropy constant $D=7.4$ meV (similar to the one reported for FeTPP crystals \cite{boyd_paramagnetic_1979}). 
A small fraction of molecules ($\sim$15$\%$) exhibit the inelastic step split into two smaller steps even in the absence of magnetic field, at V$_{s}=\pm 8.7$ and $\pm 6.5$ mV (see Fig. S2 in SM \cite{SOM}). This is probably due to small distortions of the molecular structure on the surface that causes in these cases a finite component of  transverse magnetic anisotropy  \cite{heinrich_tuning_2015}.

The above spin-model successfully predicts excitation energies and transition rates  \cite{gatteschi_molecular_2006}. However, to fully describe the spectral shape of inelastic tunneling processes spin-electron interactions must be taken into account. Electron scattering mechanisms contributing to the inelastic signal can be described by Hamiltonian terms of the form $H_{int}\simeq \mathscr{U} + J\cdot {\bf{s}}\cdot {\bf{S}}$. The exchange term $J$ describes the transfer of energy and angular momentum by electrons (with spin $\bf{s}$), and accounts for excitations of the molecular spin $\bf{S}$. The potential scattering term $\mathscr{U}$ reflects charge scattering processes by partially occupied localized states \cite{schrieffer_relation_1966,ternes_spin_2015}, and its role has typically been disregarded in magnetic systems.

\begin{figure}[ht!]
\includegraphics[width=1\textwidth]{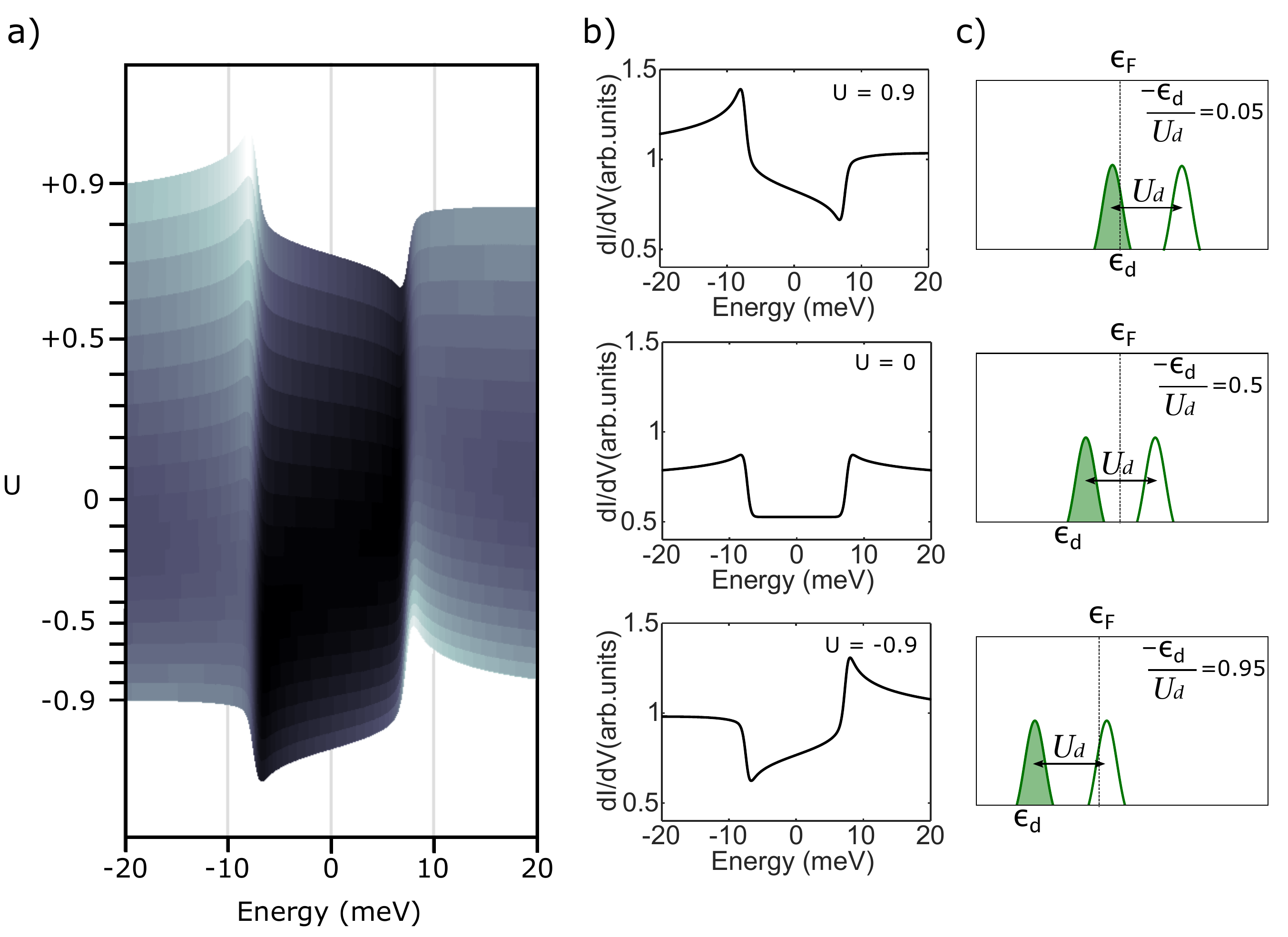}
\caption{\label{FigS5} (a) Evolution of the simulated dI/dV spectra with the ratio of potential and exchange scattering amplitudes $U=\frac{\mathscr{U}}{J}$ \cite{ternes_spin_2015}. (b) Three representative cases, namely electron-hole symmetry ($U=0$) and mix-valence regime ($U=\pm0.9$). (c) Representation of the single-impurity Anderson model parameters for the three selected cases in (b).}
\end{figure}

Most of the scattering processes contribute to the tunneling conductance with bias-symmetric components. Asymmetry in the spectra of Kondo systems has been generally attributed to Fano-like interference of the exchange-scattering channel with other tunneling paths \cite{JacobPRB2015}. While in the present case Kondo-screening can be excluded due to the large anisotropy, high-order scattering processes still apply.  In particular, it has been shown \cite{ternes_spin_2015} that the presence of a potential scattering channel can produce a peculiar bias-asymmetry in the conductance spectra. As we show in the Supplemental Material (see Fig. S6 in SM \cite{SOM}), a finite $\mathscr{U}$ amplitude leads to antisymmetric components in the $dI/dV$ spectra, arising already at biases below the excitation energy and with small logarithmic peaks/dips at the excitation  onset. As a result, the shape of the inelastic features in $dI/dV$ spectra reflect the degree of particle-hole asymmetry of the system (Fig. \ref{FigS5}).

Following this interpretation, we separate the symmetric $G_s$ and antisymmetric $G_a$ parts of the spectra in Fig. \ref{Fig2}(a)-(b) as indicated in the caption. The resulting plots (Fig. \ref{Fig2}(c)) reveal that the inelastic fraction in the symmetric component is larger over the pyrrole groups by a factor of 3 (inelastic fraction $\frac{\Delta G_s}{G_{s}}$ amounts to 0.9 and 0.3 over pyrroles and Fe, respectively), while the antisymmetric component $G_a$ has opposite sign on each site, and is 2 times larger over the Fe ion. Furthermore, the $G_a$ component shows characteristic dips and peaks at the onset of excitation, which  resemble the  antisymmetric components  of higher order terms associated to a non-zero potential scattering amplitude  (see SM \cite{SOM}), rather than other known sources of asymmetry, such a Fano-like interference with other tunneling channels.

We thus fitted the $dI/dV$ spectra of Fig. \ref{Fig2}(a) using a (second-order scattering) phenomenological model developed by Ternes \cite{ternes_spin_2015} (for fit results see Fig. S5  and Table I in SM \cite{SOM}). The stronger antisymmetric component of the curve taken over the Fe ion is reproduced using a large and positive ratio between potential and exchange scattering amplitudes, i.e. $U = \mathscr{U}/J=0.8$, whereas the opposite and weaker  antisymmetric component of the curve taken on the pyrroles is due to a smaller and negative ratio, $U = -0.4$. In Fig. \ref{FigS5}(a) we show the evolution of the asymmetry in the spectra as we sweep the parameter $U$ (see Eq. S3 in SM \cite{SOM}). The inelastic spectra shows a characteristic assymetry, that changes sign as we move from the hole ($U = 0.9$) to the particle ($U = -0.9$) mix-valence regimes, while at the particle-hole symmetry point the spectra is symmetric. Three cases are individually plotted for clarity in Fig. \ref{FigS5}(b), resembling the asymmetry found in the experiments.

The microscopic Anderson model \cite{schrieffer_relation_1966,anderson_localized_1961,ternes_spin_2015} provides a physical interpretation for the phenomenological parameter $U$ obtained from the fit to the experimental spectra. As described by the Schrieffer-Wolff transformation \cite{schrieffer_relation_1966}, $U$ depends linearly on the ratio between energy position of the singly occupied spin state and the on-site Coulomb energy $\epsilon_{d}/U_{d}$. This relation conveys information about the relative alignment of the spin energy levels with respect to the Fermi energy. In Fig. \ref{FigS5}(c) we included the representation of the orbital energy position for the three cases shown in Fig. \ref{FigS5}(b).

From the fitting of our experimental data for the more symmetric STS lineshape taken on the pyrroles, we obtain $\epsilon_d=-0.7 U_d$, which can be interpreted as a spin state close to the electron-hole symmetry point ($\epsilon_d=-0.5 U_d$), where potential scattering would be absent ($U=0$). In contrast, from spectra on the Fe site we obtain $\epsilon_d=-0.1 U_d$ a situation where the spin is close to the mixed-valence regime and the potential scattering is significant. Although both scenarios deviate from the electron-hole symmetry point, on the central Fe ion the situation is extreme and, therefore, the spectra shows a larger antisymmetric component. Even if the Anderson model does not capture the complexity of the spin density distribution of a metal-organic system, it still provides a suitable interpretation for the energy alignment of the orbitals mediating the excitation in terms of electron-hole asymmetry.


\begin{figure}[t]
	\includegraphics[width=1\columnwidth]{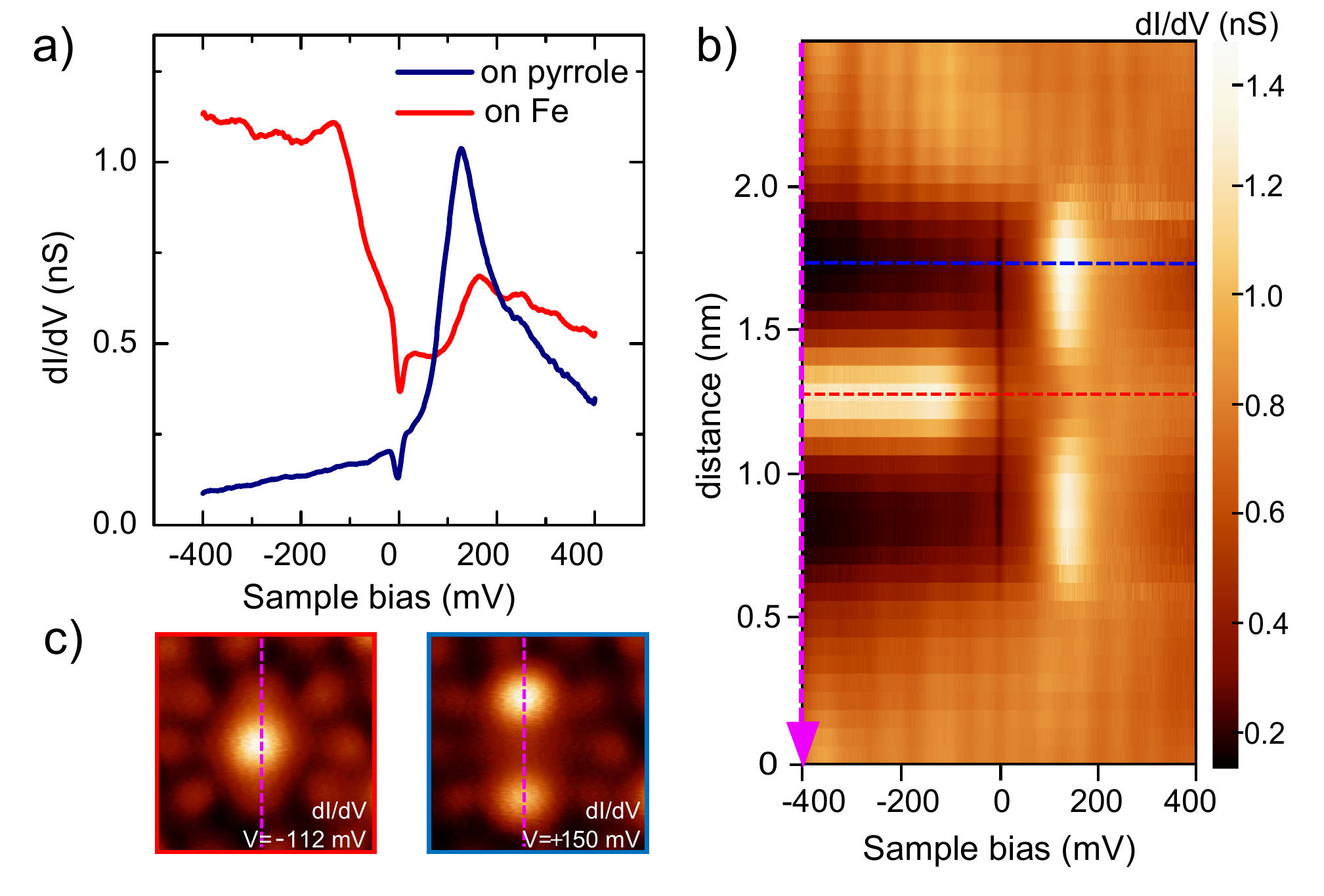}
	\caption{\label{Fig3} (a)  $dI/dV$ spectra of FeTPP measured on the two bright pyrrole groups (blue) and on the central Fe atom (red) (setpoint: $V=0.4 $ V, $I=300$ pA. Lock-in frequency 938 Hz, modulation 2 mV {\it{rms}}). The spectra is measured in a wider range as in Fig. \ref{Fig2}(a) to capture the shape and position of frontier orbitals. (b) Stacking plot of point $dI/dV$ spectra (40 curves, 2.5 nm) along the FeTPP molecule showing the spatial localization of $dI/dV$ signal along the axis determined by the two bright pyrroles.   (c) Constant height conductance maps at the energy of the resonances observed in (a).  }
\end{figure}

The energy alignment of the FeTPP frontier orbitals can be explored in spectra measured in a wider bias range. In consistency with the above-described picture, we find two electronic resonances around the Fermi level, each showing a different linewidth and spatial dependence (Fig. \ref{Fig3}(a)). While a sharp resonance appears at $V_{S}=+150$ mV
on the pyrroles, a broad state is found at negative bias over the Fe site. The $dI/dV$ stacking plot along the molecular axis (\ref{Fig3}(b)) captures these differences, and reveals an orbital pattern similar to that of the spin excitations (Fig. \ref{Fig2}(b)). The $dI/dV$ maps in Fig. \ref{Fig3}(c) further localize the positive and negative frontier orbitals along two of the four pyrroles and over the Fe center, respectively. This agrees with a molecular symmetry breaking, also observed in the spin excitation maps (Fig. \ref{Fig2}(a)). We note that the broad negative-bias resonance on the Fe ion crosses through zero bias, revealing a situation close to the mixed-valence regime, in agreement with the outcome of the Anderson model from above.

\begin{figure}[!ht]
 	\includegraphics[width=1\columnwidth]{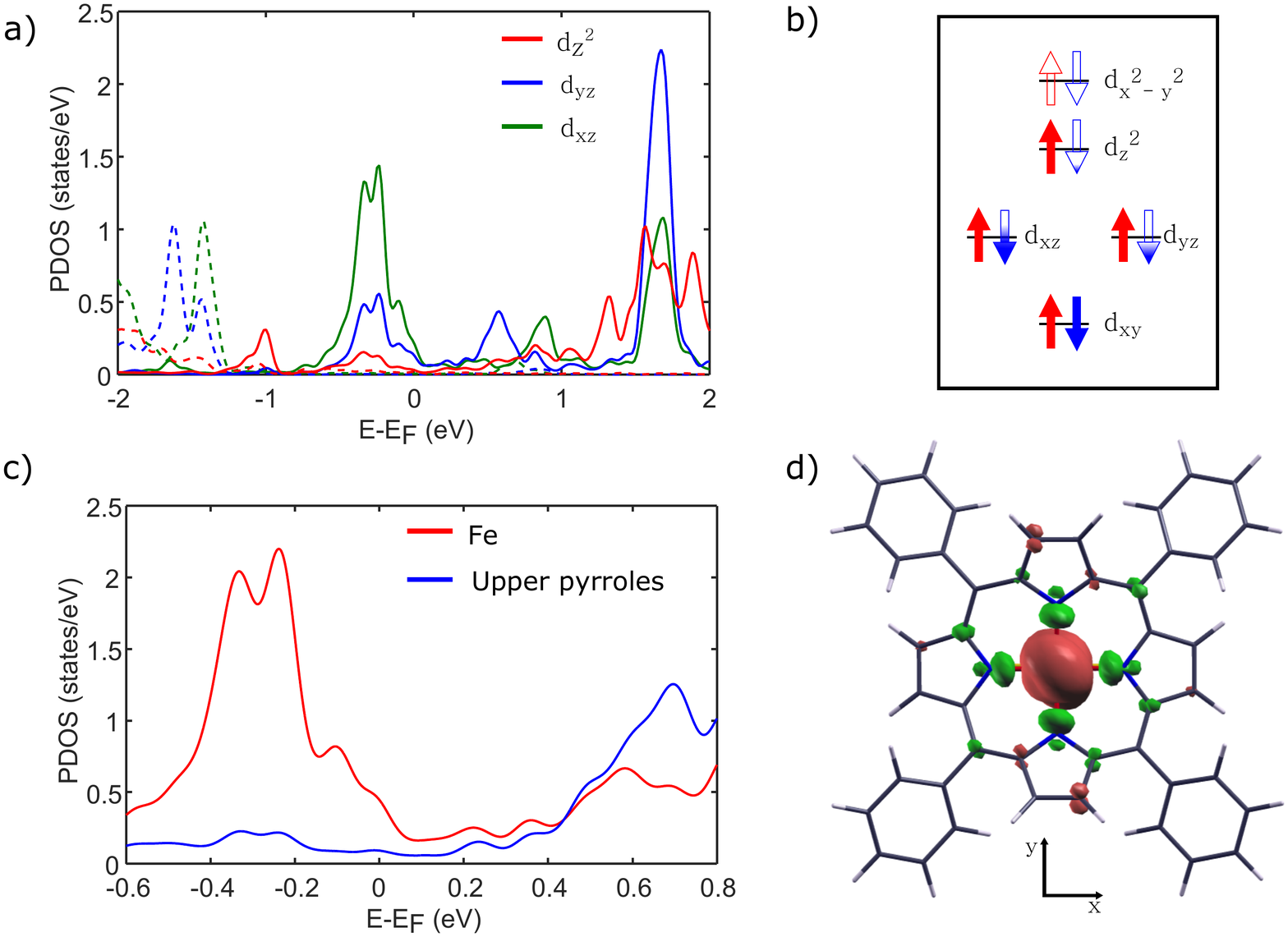}
 	\caption{\label{Fig4} (a) Spin-polarized DOS of the FeTPP/Au(111) system  projected on Fe $d_{xz}$, $d_{yz}$ and $d_{z^2}$ orbitals (dashed/solid lines represent majority/minority spin states). (b) Simplified scheme describing the occupation of  Fe 3$d$ levels for the adsorbed FeTPP (red/blue arrows indicate majority/minority spin states). The different occupation of each  state is indicated by the color filling of the corresponding arrows. (c) Total DOS of the FeTPP/Au(111) system projected on Fe states and on C and N states of the upper pyrrole moieties. (d) Spin polarization isosurface  obtained for the   FeTPP/Au(111) system by subtracting occupied states  with spin up and down (red/green surfaces describe majority/minority spin states; upper pyrroles are at the top \& bottom of the structure; the Au(111) surface has been omitted from the plot for clarity). }
 \end{figure}

In order to identify the orbital character of the frontier spectral features of FeTPP on Au(111) we analyzed its electronic configuration and spin state by means of DFT simulations based on the SIESTA code \cite{siesta}. We find that adding a Hubbard-like term with \mbox{$U_d=2$ eV} to describe the Coulomb interactions between electrons in the Fe $3d$ shell is crucial to reproduce the orbital alignment around $E_F$ observed  in the experiment \cite{Brumboiu2016}. For the free-standing molecule, our results confirm the well-known multi-configurational character of the ground state of FeTPP \cite{Vancoillie__multiconf_2011,Liao_multiconf_2007}. 
Correspondingly, we found almost degenerate solutions with quite 
different fillings of the levels associated with the metallic center and the same total spin. On the Au(111) surface the molecule adopts a distorted saddle configuration, as previously reported for porphyrins on surfaces \cite{auwarter_porphyrins_2015}. Fig. \ref{Fig4}(a) shows the density of states of FeTPP/Au(111) projected on Fe $d$ orbitals (PDOS). The FeTPP on Au(111) has a total spin $S=1$ shared between frontier  states, which have strong character on $d_{z^2}$ and $d_{\pi}$ orbitals of the Fe ion (see the sketch in Fig. \ref{Fig4}(b)). The $d_{xy}$ and $d_{x^2-y^2}$-derived states are fully occupied and empty, respectively, and largely localized due to their weak interaction with the metal substrate. The $d_{z^2}$-derived state has a net spin polarization of $0.75$ $\mu_B$ and is quite broaden as a consequence of its large hybridization with the substrate. The $d_{\pi}$-derived states appear with a broken degeneracy caused by the saddle-like distortion of the molecular backbone on the substrate. The lowest energy configuration finds $d_{yz}$ orbital with larger spin density, while the $d_{xz}$ is almost completely occupied  and, thus, exhibits a substantially smaller spin polarization. Namely, the spin polarization coming from the $d_{\pi}$ orbitals is dominated by the contribution from $d_{yz}$. The computed PDOS  thus pictures the multi-orbital character of the spin polarization of this system.

The spatial spin excitation pattern observed in the experiment is a consequence of different molecular states involved in the tunneling through the molecule. In Fig. \ref{Fig4}(c) we compare the PDOS on Fe orbitals and on the upper-pyrrole groups. 
The former resemble the $dI/dV$ spectrum at the center of the molecule (Fig. \ref{Fig3}(a)) with larger DOS at the negative part of the energy spectrum  plus some weight above E$_F$. This proves that the tunneling transmission on the Fe ion is governed by Fe $d$-resonances, probably via the broad $d_{z^2}$ components, which extend further into the vacuum.   The larger overlap of this Fe-$d$ manifold with $E_F$, closer to a mixed-valence configuration, is the main cause of the larger asymmetry of the spin excitation features observed over the Fe site.  In contrast, the PDOS on the upper-pyrrole atoms exhibit only  a  resonance at positive energy, which  correlates with the $dI/dV$ peak measured over the protruding pyrrole groups (Fig. \ref{Fig3}(a)). The resonance is composed of  pyrrole states  and empty $d_{yz}$ component of the Fe ion, having a strong hybrid Fe-ligand character. As a consequence, this resonance becomes weakly spin-polarized (Fig. \ref{Fig4}(d)).  
 Therefore, tunneling through pyrrole-Fe hybrid  states can effectively excite the spin of the Fe ion \cite{Strozecka2012}, whereas the smaller overlap of this resonance with E$_F$ agrees with the more symmetric excitation lineshape observed in the experiment.


In summary, we have demonstrated that the spin excitation of a S=1 metal-organic molecule can be selectively excited through two different inelastic tunneling channels. Every channel is mediated by a molecular state with different spatial extension and produces inelastic features with a characteristic asymmetry. We have shown that the asymmetry is a consequence of the spin-polarized molecular state being out of the particle-hole symmetric case. Thus, interpretation of the inelastic spectra in terms of a phenomenological scattering model can be used to detect the alignment of spin-hosting states in molecular magnets and to infer their proximity to a mix-valence configuration.



We thank Nicol\'as Lorente and Thomas Frederiksen for stimulating discussions. This work has been funded by the COST 15128 Molecular Spintronics project, by Marie Curie IF ARTE, by the Spanish Ministerio de Econom\'ia y Competitividad (MINECO) through the cooperative grant No. MAT2016-78293 and grant No. FIS2016-75862-P, and by the Basque Government (Dep. Industry, Grant PI-2015-1-42, Dep. Education, Grant PI-2016-1-27 and Grant IT-756-13), the EU project PAMS (610446), and the European Regional Development Fund (ERDF).

\section*{Methods}

Our experiments were performed in a low temperature Scanning Tunneling Microscope (STM) with a base temperature of 1.2 K under UHV conditions (JT-STM by SPECS GmbH). We thermally sublimated 5,10,15,20-Tetraphenyl-porphine iron(III) chloride (FeTPP-Cl) molecules on the clean Au(111) substrate at room temperature. Differential conductance ($dI/dV$) measurements were acquired using lock-in amplifier technique. Analysis of STM and STS data was performed with the WSxM \cite{horcas_wsxm:_2007} and SpectraFox \cite{spectrafox} software packages.

\bibliography{porphyrins}

\end{document}